\documentclass[showpacs,preprint,amssymb]{revtex4-1}
\usepackage{graphicx}
\usepackage{dcolumn}
\usepackage{bm}
\usepackage{amsmath}
\usepackage{amsfonts}
\usepackage{amssymb}
\usepackage{ulem}
\usepackage{color}

\begin{document}
\title{Mechanically-Induced Transport Switching Effect in Graphene-based
Nanojunctions}

\author{T. Kawai$^1$}
\author{M. Poetschke$^2$}
\author{Y. Miyamoto$^1$}
\author{C.G. Rocha$^2$}
\author{S. Roche$^{2,3,4}$}
\author{G. Cuniberti$^{2,5}$}
\affiliation{$^1$ Green Innovation Research. Laboratoriess, NEC Corp., 34
Miyukigaoka, Tsukuba, Ibaraki, Japan}
\affiliation{$^2$ Institute for Materials Science and Max Bergmann Center
of Biomaterials, Dresden University of Technology, D-01062 Dresden, Germany}
\affiliation{$^3$ Centre d' ~Investigac\'\i o en Nanoci\`encia i Nanotecnologia (ICN-CSIC), UAB Campus, E-08193 Bellaterra, Spain}
\affiliation{$^4$ Instituci\'o Catalana de Recerca i Estudis Avancats (ICREA), 
E-08100 Barcelona, Spain}
\affiliation{$^5$ Division of IT Convergence Engineering, POSTECH, Pohang 790-784, Republic of Korea}

\date{\today}

\begin{abstract}
We report a theoretical study suggesting a novel type of electronic
switching effect, driven by the geometrical reconstruction of nanoscale 
graphene-based junctions. We considered junction
structures which have alternative metastable configurations transformed 
by rotations of local carbon dimers.
The use of external mechanical strain allows a control of the energy
barrier heights of the potential
profiles and also changes the reaction character from endothermic to
exothermic or vice-versa. The reshaping
 of the atomic details of the junction encode binary electronic ON or OFF
states, with ON/OFF transmission ratio
 that can reach up to 10$^4$-10$^5$. Our results suggest the possibility
to design modern logical switching devices or mechanophore sensors, monitored by
mechanical strain and structural rearrangements.
\end{abstract}

\pacs{...}
\maketitle

Current strategies for miniaturization of logic and memory devices
envision the ultimate limit which is the atomic scale control.
Several experimental groups have demonstrated the capability
of controlling the current flow through single molecular junctions
~\cite{Chen1999,Scheer1998,Xie2006,Nozaki2009,Patolsky2004} highlighting
promising applications in realms such as nanotechnology and medicine. Outstanding
switching properties  have been recently revealed in atomic-scale metallic contacts
disposed in a three-terminal architecture where the conductance was
mechanically switched  between ``ON'' and ``OFF'' states at room temperature~\cite{Xie2008}.
Other attractive materials such as photochromic organic molecules
have been widely used for designing LCD displays and liquid crystals since
their transport features can be driven by light irradiation~\cite{Brakemann2010} 
or current pulses~\cite{Zhang2004}. In the quest of designing efficient
atomic-scale switching devices, graphene-based materials deserve
special attention due to a peculiar one-atom-thick planar geometry
combined with exceptional electronic~\cite{Tan2007}, mechanical and thermal
 properties~\cite{Novoselov2004,Castro2009,Janina2008}. As a result, a wealth of novel phenomena is steadily unveiled
in different fields, including spintronics~\cite{Son2006}, ac
transport~\cite{Rocha2010,Prada2009,Zhu2009}, or
thermoelectrics~\cite{Zuev2009}.

However, the use of graphene for designing logical switching applications
is severely limited by the absence of energy gap. Using advanced lithographic
techniques~\cite{Datta2008,Lemme2009},
the fabrication of graphene nanoribbons (GNRs) and band gap engineering
have paved the way towards more efficient switching devices. 
Patterned monolayer graphene nanoconstriction-based field effect transistors were 
successfully synthesized, confirming that structural confinement can
induce energy gap opening and a sharp increase in the ON/OFF ratio ($\sim$10$^4$)~\cite{Lu2010}. 
GNRs-based electromechanical and
electromagnetic switches were also designed, evidencing the possibility of switching principles of
different physical natures~\cite{Milaninia2009,Dragoman2009}.
Recently, an efficient and reversible current-induced switching mechanism
was reported in large area-graphene based devices, allowing for the design of
nonvolatile memory elements~\cite{Standley2008}.
The origin of this switching phenomenon was tentatively related to the
formation and breaking of carbon atomic chains bridging two sections of the sample. 
 Additional control parameters can be considered to modulate in a reversible fashion the binary switching properties of molecular junctions. For instance,
recent studies have confirmed that mechanical strain can effectively monitor
the electronic structure and transport response of carbon-based materials
~\cite{Quek2009,Ni2009,Poetschke2010,Mohiuddin2009,Pereira2009}. In particular for
graphene that are gapless materials, the use of mechanical forces as control
parameters can establish important strategies for envisioning band gap engineering in graphene
structures.

Despite all these efforts to develop
and monitored switching devices, the understanding of
intrinsic operating mechanisms
and their relation with the atomic-scale features are open issues
demanding for advanced theoretical investigation.
In particular, there is a need to investigate the deep connection between
reversible modifications occurring at the
 atomic-scale and the resulting electronic behavior of devices exposed to
external perturbations~\cite{Agapito2007,Yin2009,Hashimoto2004,Yakobson2000}.
In this Letter, the inherent switching mechanism of physical reactions
involving graphene etched junctions is dynamically probed using theoretical
simulations. Our goal is to identify
optimum physical conditions for tuning the conductance between binary
switching states. 
A physical reaction is simulated by examining the potential profile on the 
coordination axis of the rotating carbon dimers under applied 
mechanical stress needed for improving the switching features of the device.
The associated activation energy for the reaction
and its ground state are determined by the derivation of minimum energy
paths (MEP). Mechanical strain is subsequently used to modulate the
energy onsets and barrier heights which characterize the structural
transition. Regarding the transport response of the system, we establish clear
thresholds between ON and OFF states as
we monitor their electronic transmission along the whole reaction path.
Prominent switching behaviour can be achieved depending on the atomic
widths of the nanostructures and amount of applied strain.
Our findings ensure a promising way for designing ultimate
molecular switching devices based on strained graphene.
We also extend graphene's applicability
to another scientific frontier which is the production of
 mechanophore sensors based on etched graphene samples. Such special
materials consist of
systems where their reaction path depends on the strength and nature of
the applied force~\cite{Arino2010}.

\begin{figure}[ptbh]
\begin{center}
\leavevmode
\includegraphics[scale=0.5]{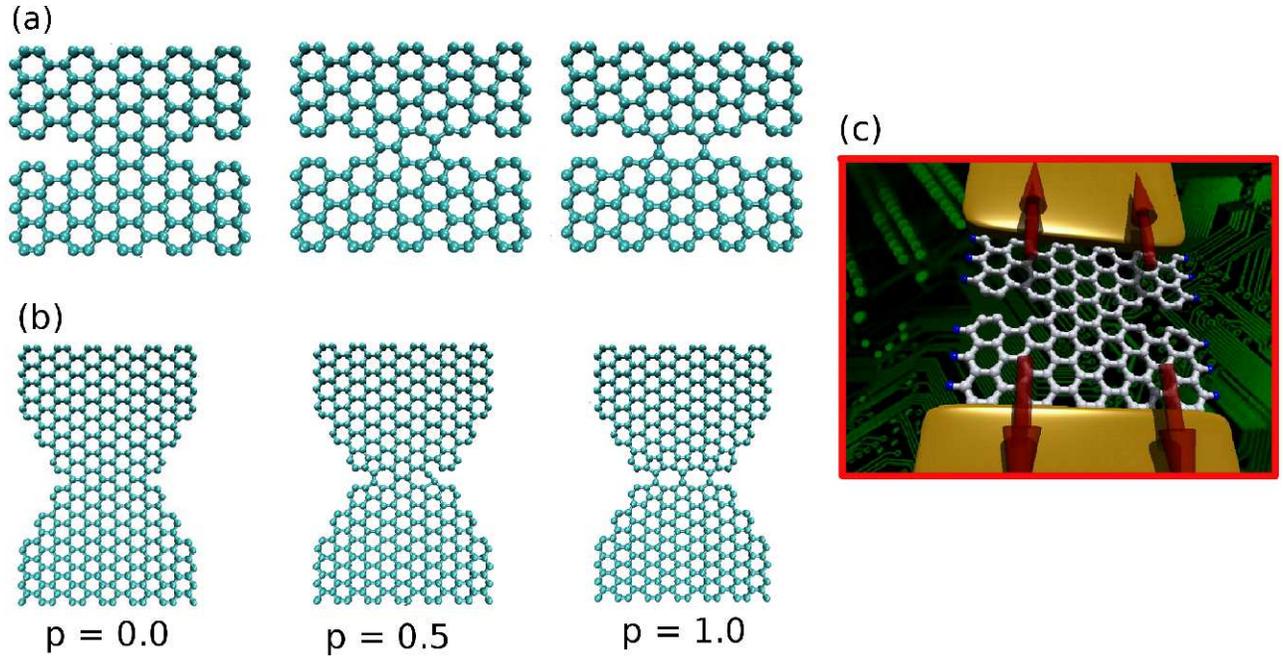}
\caption{The corresponding (ball and stick)
atomic configurations for initial ($p=0$),
 intermediate ($p=0.5$) and final states ($p=1.0$)
for a (a) bi- and (b) tri-switching structures.
Panel (c): Ball and stick model of a graphene-based junctions
under uniaxial mechanical forces (arrows).}
\label{fig1}
\end{center}
\end{figure}

\textit{Minimum energy path results:}
In what follows, we performed systematic analysis of transport properties
and mechanical activity of two different
graphene constrictions where carbon dimers
bridge the gap between two semi-infinite graphene ribbons. Here we refer
as a bi- (tri-) switch
configuration, the constriction with two (three) C-C bonds linking
the graphene segments.
A minimum energy path which resolves the
system undergoing a structural rearrangement of atoms
can be tracked assuming two metastable configurations:
a graphene nanojunction made by pure hexagons (initial state) and a defective
constriction composed of pentagon-heptagon pairs (final state). The
variable $p$ is associated to the normalized reaction coordinate
where $p=0$ corresponds to the initial anchor state and $p=1.0$ defines
the ending of the reaction.
Ball and stick schematic pictures for initial, intermediate and final states
are displayed in Fig.~\ref{fig1}, panels (a) and (b). The physical reaction consists of
rotating the C-C bonds located on the ``neck'' of the initial hexagonal
junction by 90$^\circ$, generating rings of Stone-Wales defects in the final
configuration. After the rotation, we observe that the optimal
lattice constant along the axial direction suffers a small expansion 
of $\sim$ 0.7 \AA, characterizing an endothermic-like reaction.
The transition route linking these two configurations states
can be located using nudged elastic band (NEB) method \cite{Henkelman2000}.
The method gathers the MEP by constructing a set of replicas
of the system between the initial and final states. A spring
interaction between consecutive images is included to ensure the
continuity of the path. Optimization procedures that minimize the forces
acting on the images are implemented to obtain the MEP. 

\begin{figure}[ptbh]
\begin{center}
\leavevmode
\includegraphics[scale=0.4]{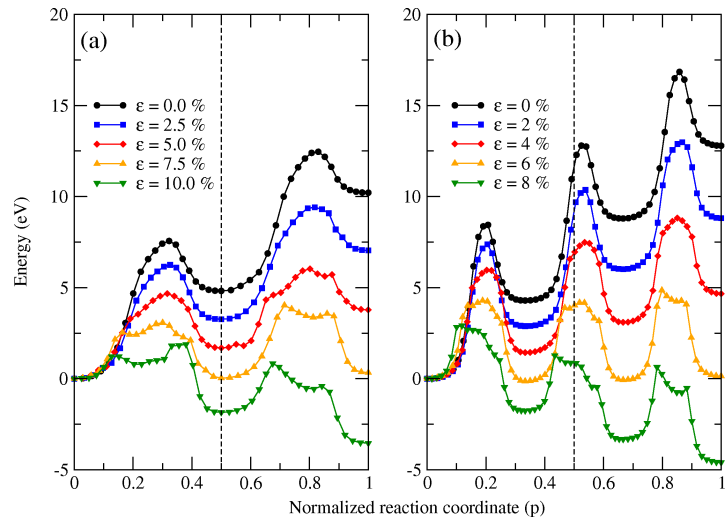}
\caption{Minimum energy paths calculated for (a) bi- and (b) tri-
switching structures for different values of mechanical strain. The 
vertical dotted lines in the graphs denote the corresponding 
positions of intermediate states ($p=0.5$) in the reaction coordinate.
Reference energy for the initial configuration
was set to 0 eV.}
\label{fig2}
\end{center}
\end{figure}

Our results obtained
for the two switching systems are shown in Fig.~\ref{fig2}.~
The total energy variation along the whole trajectory are mapped as a
function of the reaction coordinate ($p$) and the reference energy for the initial state is
set to 0 eV. The structures are also submitted to uniaxial mechanical forces 
under DFTB frame~\cite{Xu92} and we
demonstrate the ability of controlling the MEP profiles by applying such perturbation.
A schematic representation of the applied forces is shown in
Fig.~\ref{fig1} (c). 
We applied mechanical strain by pulling away the atoms
located at upper and lower edges at incremental values and further fixing them.
The atomic geometries except for the edge constrained atoms are subsequently relaxed under the 
conditions of NEB method.
The equivalent mechanical energy resulting from the stretching is defined as
$E_M=-\Delta \vec{L}\cdot \vec{F}$, being
$\vec{F}$ the external force and $\Delta L = L-L_0$, with $L$ and $L_0$ the
deformed and initial equilibrium lattice constant parameters along the
axial direction, respectively. The strain is calculated as $\epsilon=\Delta
L/L_0$.

For unstrained systems, the ground state for both switching systems is
the initial hexagonal junction. For bi-switching configuration, two humps can be
observed along the reaction path while three others appear for tri-switching
arrangement. Each one of those humps is associated to the amount of energy required to
rotate one C-C bond by 90$^\circ$ in the junction.
After the rotation, we observe that the optimal lattice constant along the axial direction
suffers a small expansion of $\sim 3$\%, characterizing
an endothermic reaction. The energy barrier heights for the
complete transition are estimated as 12.4 eV and
 16.8 eV for bi-switching and tri-switching structures, respectively.
The barrier heights as well as the energies of initial/final
configurations can be strongly modulated
with the aid of mechanical strain.
From the pictures, we can see that the barrier heights are considerably
reduced as stretching takes place. At tensile strains of 7.5\% (6\%) for bi- (tri-)
switching structure, the barrier heights for
both structures decrease from 12.4 eV (16.8 eV) to 4.0 eV (4.8 eV), 
and the anchor states are set almost evenly in energy.
As the mechanical strain continues to increase, the ground state is
switched from the hexagonal
junction to the 5-7 defective configuration and the reaction becomes
exothermic-like. The formation of 5-7 defect lines reduces the number of the
bonds connecting the two graphene segments and this dwindles the energy loss
in response to the elongation.

Up to this point, we evidenced that the physical reactions involving graphene
junctions under strains of 6\%-7.5\% are energetically favorable to be
triggered since the total energies of initial and final states are comparable and
the barrier heights are considerably reduced.
We now explore if such strained systems reveal optimum switching behaviour
through electronic structure analysis. This issue is investigated by
following how the electronic
structure and transport responses change along the reaction coordinate path.
The efficiency of standard logical switches can usually be measured
through the
magnitude of ON/OFF transmission rates, controlled by gate voltages. In
our case, the reaction coordinate
that resolves the dynamic trajectory between the anchor states, acts as an
additional switching parameter.
In this sense, prominent switching responses can also be dynamically tuned
as the system evolves along the MEP.
An efficient dynamic switch is characterized by finding pronounced
differences
between the transmission values taken from the initial and final
configurations.
To represent the electronic structure of both extreme states, we adopted
self-consistent-charge Density Functional based tight binding
theory (DFTB)~\cite{Elstner1996,Seifert1996} and the conductance was
subsequently calculated using Landauer formula written in terms of
Green's functions~\cite{Nardelli1999}.

\begin{figure}[ptbh]
\begin{center}
\leavevmode
\includegraphics[scale=0.30]{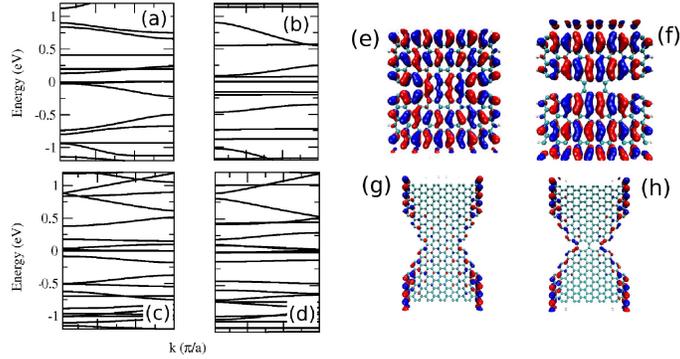}
\caption{Bandstructures obtained for
initial [(a) and (c)] and final [(b) and (d)] states of a bi- (upper panels)
and tri- (lower panels) switching graphene junctions under strain (7.5\%
and 6\%, respectively). Besides is shown
the electronic wave functions
calculated at $\Gamma$ point at the energies (e) -0.74 eV, (f) -0.72 eV,
(g) -0.02 eV and (h) -0.01 eV.
Red and blue colors correspond to positive and negative coefficients,
respectively.}
\label{fig3}
\end{center}
\end{figure}

Figure ~\ref{fig3} depicts the energy relation dispersions calculated for
initial [(a) and (c) panels] and final states [(b) and (d) panels] for bi-switching (upper panels)
and tri-switching (lower panels) systems. The Fermi energy is set at 0 eV.
The atomic arrangement along the edges of both junctions follows a zigzag
shape and hydrogen atoms are used to saturate dangling bonds situated outside
the constriction domain. The non-saturated dangling bonds give raise to
flat bands at the  Fermi level. This can be confirmed through
the electronic wave function plots displayed on the lower 
panels (g) and (h) for the tri-switching example. The wave functions are
calculated at the $\Gamma$ point for energies around $E_F=0.0$ eV.
In fact, the electronic distribution is highly concentrated on the edges.
Such bands are rather robust and do not respond to the structural
modification of rotating carbon dimers. This indicates that
favorable switching features cannot be achieved at $E_F = 0.0$ eV for both
nanojunctions.

For the bi-switching arrangement, outstanding switching
properties can be highlighted in the energy range of $-0.5 < E <-0.1$ eV.
The subband dispersion within this range shown on panel (a) is
considerably reduced
as the bonds rotate, characterizing a potential transition between ON
[panel (a)] and OFF [panel (b)] states.
To rationalize these findings, the electronic wave functions
calculated at the $\Gamma$ point are shown for particular energies in the bi-switching
example [(e) and (f) lower panels].
We observe a highly delocalized spatial distribution of conduction channels
for the initial frame whereas two disconnected orbital patterns are obtained
at the ending of the reaction. Rotating the horizontal C-C bonds of the
hexagonal bi-switching junction disrupts the superposition of the molecular orbitals on the
constriction. The system behaves, then, as two isolated semi-infinite graphene sections
and the electronic channel is blocked.
On the other hand, the electronic structure of tri-switching arrangement
seems to exhibit weaker switching behaviour.
Comparing the energy relations displayed on panels (c) and (d), one can see
a significant resemblance within the whole spectrum
which means that the structural reaction does not cause
significant impact on the electronic response of such junctions.

\begin{figure}[ptbh]
\begin{center}
\leavevmode
\includegraphics[scale=0.2]{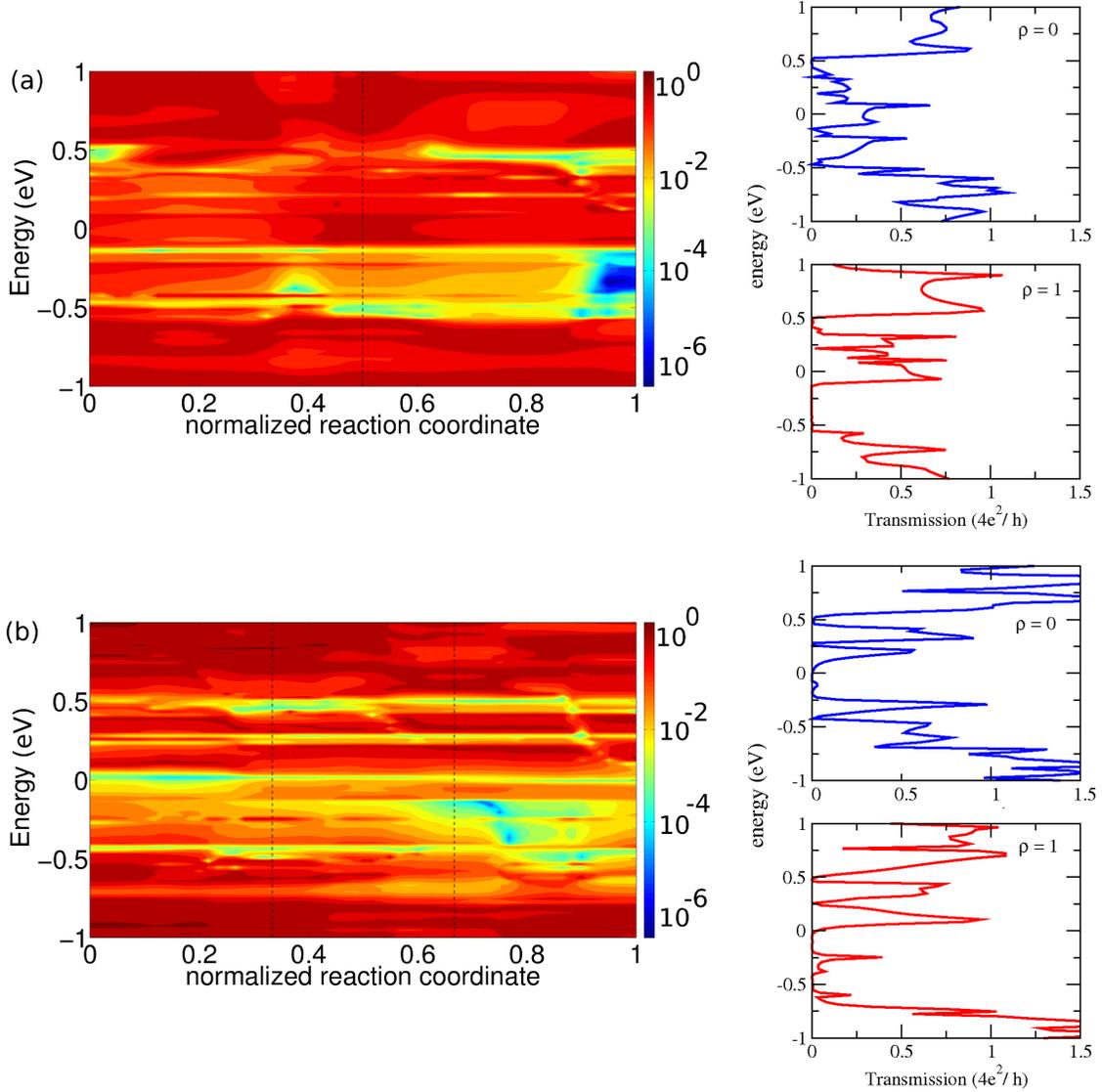}
\caption{Decadic logarithm of the transmission
as a function of reaction coordinate and Fermi energy for a (a) bi- and
(b) tri-switching structure
under mechanical strain (7.5\% and 6.0\%, respectively). Panels beside
the contour plots display the transmission curves as a function of energy
at initial ($\rho = 0$) and final ($\rho = 1$) reaction coordinates for
(upper panels) bi- and (bottom panels) tri-switching configurations.}
\label{fig4}
\end{center}
\end{figure}

The precise evaluation of the ON/OFF transmission rates is obtained
through Fig.~\ref{fig4} which shows the conductance contour plots
as a function of the Fermi energy and reaction coordinates obtained
for both strained samples.
The results reveal once more the excellent binary switching
behaviour of bi-switching structure in contrast to tri-switching case. The gradient color
located on
the energy range of $-0.5 < E <-0.1$ eV unveils a gradual
exchange between ON (red color) and OFF (blue color) states
with respect to the reaction coordinate path. The maximum transmission
rate is determined as 10$^4$-10$^5$. Assuming that the final state can be
stabilized, the transition between high and low transmission
states can also be induced through the use of some external gate voltages.
Differently,
tri-switching configuration manifests a robust transport response
with minor changes appearing in the same energy range. Lower
transmission values are evidenced only for a few intermediate
reaction frames indicating that the weaker switching capability
of such system can be better manipulated via gate voltages.

\textit{Conclusion:} We reported the manifestation of binary switching
behaviour on graphene nanojunctions that experience a physical reaction
capable of rearranging locally the atomic structure of the systems.
The junctions are also exposed to uniaxial mechanical stretching
which tunes the physical characteristics of the reaction such as
energy barrier heights and ground state. 
As the reaction occurs, we observe
that narrower junctions reveal promising
ON/OFF transmission rate reaching values up to 10$^4$-10$^5$ which is in agreement with current
experimental measurements performed in graphene nanoconstrictions\cite{Lu2010}. 
Our results also support the interpretation of recent experiments carried 
on atomic quantum transistors where self-stabilizing contact reconstruction can be mechanically
driven~\cite{Xie2008}.

\textit{Acknowledgments.}
This work was partially funded by Alexander von Humboldt
Foundation, the European Union, the Free State of Saxony
(SAB project number A2-13996/2379) and the NANOSIM-GRAPHENE
Project No. ANR-09-NANO-016-01 and the South Korean Ministry
of Education, Science, and Technology Program, Project WCU
ITCE No. R31-2008-000-10100-0.


\begin{thebibliography}{99}

\bibitem{Chen1999} J. Chen, {\it et al.}, Science {\bf 286}, 1550 (1999).

\bibitem{Scheer1998} E. Scheer, {\it et al.}, Nature {\bf 394}, 154 (1998).

\bibitem{Xie2006} F.-Q. Xie, Ch. Obermair and Th. Schimmel,
NATO Sci. Series II: Math., Phys. and Chem. {\bf 233}, 153 (2006).

\bibitem{Nozaki2009} D. Nozaki and G. Cuniberti, Nano Res. {\bf 2},
648 (2009).

\bibitem{Patolsky2004} F. Patolsky, {\it et al.}, Proc. Natl.
Acad. Sci. USA {\bf 101}, 14017 (2004).

\bibitem{Xie2008} F.-Q. Xie, {\it et al.}, Nano Lett. {\bf 8},
4493 (2008).

\bibitem{Brakemann2010} T. Brakemann, {\it et al.}, J. Bio. chem.
{\bf 285}, 14603 (2010).

\bibitem{Zhang2004} C. Zhang, {\it et al.}, Phys. Rev. Lett. {\bf 92},
158301 (2004).


\bibitem{Tan2007} Y. W. Tan, {\it et al.}~
S. Adam, E. H.
Phys. Rev. Lett. {\bf 99}, 246803 (2007); A. Lherbier, B. Biel, Y.-M.
Niquet, and S. Roche, ibid. 100, 036803 (2008).

\bibitem{Novoselov2004} K.S. Novoselov, {\it et al.}~
Morozov, D. Jiang, Y.
Science {\bf 306}, 666 (2004).

\bibitem{Castro2009} A.H. Castro Neto, {\it et al.}, Rev. Mod. Phys. {\bf
81}, 109 (2009); M. Lemme, Sol. Stat. Phenom. {\bf 156-158}, 499 (2010). 

\bibitem{Janina2008} J. Zimmermann, P. Pavone, and G. Cuniberti,
Phys. Rev. B {\bf 78}, 045410 (2008).

\bibitem{Son2006} Y.-W. Son {\it et al.}, Nature (London) {\bf 444}, 347
(2006); S. Lakshmi,
S. Roche, and G. Cuniberti, Phys. Rev. B {\bf 80}, 193404 (2009); 

\bibitem{Rocha2010} C. G. Rocha, L. Foa Torres, and G. Cuniberti, Phys. Rev.
B {\bf 81},
115435 (2010).

\bibitem{Prada2009} E. Prada, P. San-jose, and H. Schomerus, Phys. Rev. B
{\bf 80},
245414 (2009).

\bibitem{Zhu2009} R. Zhu, and H. Chen, Appl. Phys. Lett. {\bf 95}, 122111
(2009).

\bibitem{Zuev2009} Y.M. Zuev, W. Chang, and P. Kim, Phys. Rev. Lett. {\bf
102},
096807 (2009); 

\bibitem{Datta2008} S.S. Datta, D.R. Strachan, S.M. Khamis, and A.T.C.
Johnson, Nano Lett. {\bf 8} 1912 (2008).

\bibitem{Lemme2009} M.C. Lemme, {\it et al.}, ACS Nano {\bf 3}, 2674 (2009).

\bibitem{Lu2010} Y. Lu, B. Goldsmith, D.R. Strachan, J.H. Lim, Z. Luo, A.T.C. Johnson,
Small {\bf 6}, 2748 (2010).

\bibitem{Milaninia2009} K.M. Milaninia, M.A. Baldo, A. Reina,
and J. Kong, Appl. Phys. Lett. {\bf 95}, 183105 (2009).

\bibitem{Dragoman2009} M. Dragoman, D. Dragoman, F. Coccetti,
R. Plana, and A. A. Muller, J. Appl. Phys. {\bf 105},
054309 (2009).

\bibitem{Standley2008} B. Standley, {\it et al.}, Nano Lett. {\bf 8},
3345 (2008).

\bibitem{Quek2009} S. Y. Quek, {\it et al.}, Nat. Nanotech. {\bf 4}, 230
(2009)

\bibitem{Ni2009} Z.H. Ni, T. Yu, Y.H. Lu, Y.Y. Wang, Y.P. Feng,
and Z.X. Shen, ACS Nano {\bf 3}, 483 (2009)

\bibitem{Poetschke2010} M. Poetschke, C.G. Rocha, L.E.F. Foa Torres, S.
Roche, and G. Cuniberti, Phys. Rev. B {\bf 81},
193404 (2010); 

\bibitem{Mohiuddin2009} T. M. G. Mohiuddin, {\it et al.}, Phys. Rev. B
{\bf 79}, 205433 (2009).

\bibitem{Pereira2009} V. M. Pereira, and A. H. Castro Neto, Phys. Rev. Lett.
{\bf 103}, 046801 (2009); V. M. Pereira, N. M. R. Peres, and A. H. Castro Neto,
Phys. Rev. B {\bf 80} 045401 (2009).

\bibitem{Agapito2007} L. A. Agapito, and H.-P. Cheng, J. Phys. Chem. C
{\bf 111}, 14266 (2007).

\bibitem{Yin2009} C. Yin, {\it et al.}, J. Chem. Phys. {\bf 131},
234706 (2009).

\bibitem{Hashimoto2004} A. Hashimoto, K. Suenaga, A. Gloter, 
K. Urita, and S. Iijima, Nature {\bf 430}, 870 (2004).

\bibitem{Yakobson2000} B. I. Yakobson, G. Samsonidze, and G. G. Samsonidze,
Carbon {\bf 38}, 1675 (2000).

\bibitem{Arino2010} J.R.-Arino, M. Shiga, and D. Marx, J. Am. Chem.
Soc. {\bf 132}, 10609 (2010)

\bibitem{Henkelman2000} G. Henkelman, B.P. Uberuaga, and H. J\'onsson,
J. Chem. Phys. {\bf 113}, 9901 (2000)

\bibitem{Xu92} C.H. Xu, C.Z. Wang, C.T. Chan, and K.M. Ho, J. Phys. Condens. Matter {\bf 4},
6047 (1992)

\bibitem{Elstner1996} M. Elstner, {\it et al.}, Phys. Rev. B  {\bf 58},
7260 (1998)

\bibitem{Seifert1996} G. Seifert, D. Porezag, and T. Frauenheim,
Int. J. Quantum Chemistry {\bf 58}, 185 (1996)


\bibitem{Nardelli1999} M. Buongiorno Nardelli, Phys. Rev. B {\bf 60}, 7828 (1999)

\end{thebibliography}
\end{document}